\documentclass[useAMS,usenatbib]{mn2e}
\usepackage{graphicx}
\usepackage{amssymb}

\title[]{Gravitational wave background from sub-luminous GRBs: prospects for second and third generation detectors}

\author[]{E. Howell$^{1}$\thanks{E-mail:email@address}, T. Regimbau$^{2}$, A. Corsi$^{3}$, D. Coward$^{1}$ and R. Burman$^{1}$\\
$^{1}$School of Physics, University of Western Australia, Crawley WA 6009, Australia\\
$^{2}$Dpt. ARTEMIS, Observatoire de la C$\hat{o}$te d'Azur, BP 429 06304 Nice, France\\
$^{3}$ 18-34 LIGO Laboratory, California Institute of Technology, Pasadena CA 91125, USA\\}

\begin{document}

\pagerange{\pageref{firstpage}--\pageref{lastpage}} \pubyear{2002}

\maketitle

\label{firstpage}

\begin{abstract}
We assess the detection prospects of a gravitational wave background associated with sub-luminous gamma-ray bursts (SL-GRBs). We assume that the central engines of a significant proportion of these bursts are provided by newly born magnetars and consider two plausible GW emission mechanisms. Firstly, the deformation-induced triaxial GW emission from a newly born magnetar. Secondly, the onset of a secular bar-mode instability, associated with the long lived plateau observed in the X-ray afterglows of many gamma-ray bursts \citep{Corsi:2009}. With regards to detectability, we find that the onset of a secular instability is the most optimistic scenario: under the hypothesis that SL-GRBs associated with secularly unstable magnetars occur at a rate of $(48-80)\,\mathrm{Gpc}^{-3}\mathrm{yr}^{-1}$ or greater, cross-correlation of data from two Einstein Telescopes (ETs) could detect the GW background associated to this signal with a signal-to-noise ratio of 3 or greater after 1 year of observation. Assuming neutron star spindown results purely from triaxial GW emissions, we find that rates of around $(130-350)\,\mathrm{Gpc}^{-3}\mathrm{yr}^{-1}$ will be required by ET to detect the resulting GW background. We show that a background signal from secular instabilities could potentially mask a primordial GW background signal in the frequency range where ET is most sensitive. Finally, we show how accounting for cosmic metallicity evolution can increase the predicted signal-to-noise ratio for background signals associated with SL-GRBs.
\end{abstract}


\graphicspath{{./}{MagnetarFigures_1/}}

\linespread{1.0}

\section{Introduction}

The two closest recorded gamma-ray bursts (GRBs), GRB 980425 (36 Mpc) and GRB 060218 (145 Mpc), along with GRB 031203, associated with a host galaxy at $\sim 480$ Mpc \citep{feng_09}, make up a sub-class of long duration GRBs\footnote{Hereafter, we refer to long GRBs as those having a $T_{90}$ duration $>$ 2\,s in agreement with the traditional classification by \citet{Kouveliotou_1993}} (LGRBs) known as sub-luminous GRBs (SL-GRBs) \citep{cobb_06,Liang_07,guettaDellaValle_07,Virgili_09}. This class of GRB have isotropic equivalent $\gamma$-ray energy emissions typically several orders of magnitude below those of standard long-duration GRBs  \citep{murase_06,guettaDellaValle_07,Imerito_08} suggesting that they could form a unique population of bursts.

Observations have confirmed that at least some LGRBs are associated with the deaths of massive stars \citep{WB_06}. One scenario for the generation of LGRBs is described by the collapsar model \citep{woosley_93,MacFadyen_2001}. In this model, the inner part of a progenitor star (a Wolf-Rayet, WR, star) collapses, via a Type Ib/c supernova, forming a rapidly rotating black hole. High angular momentum enables the infalling matter to form an accretion disk, which in turn powers an ultra-relativistic jet (a ``fireball") that blasts through the stellar envelope \citep{MeszarosRees_92,woosley_93,SariNarayanPiran_98}. A number of authors have suggested that, at least in some cases, GRB explosions may end in the formation of a highly magnetized neutron star (NS), i.e. a magnetar rather than a black hole \citep{Usov:1992zd,DuncanThomson_92,DaiLu_1998,Nakar:2007yr,Bucciantini_2009,Zhang_09}. Additional support for this scenario has come from detailed modeling of the spectra and light curve of SN 2006aj (Type Ic), associated with GRB 060218. This analysis suggested that the explosion energy and ejected mass originated from a progenitor star with a zero-age main sequence mass of $\sim 20 \mathrm{M}_{\odot}$, implying the birth of a NS rather than a black hole \citep{mazzali_06,sodoburg_06_LLGRBRate_06}. The fact that this burst was of the SL-GRB class suggests that a proportion of such bursts may well be powered by magnetars \citep{toma_GRB060218,murase_06}.

The magnetar scenario for GRBs has been invoked to explain recent observations by the \emph{Swift} satellite \footnote{http://heasarc.gsfc.nasa.gov/docs/swift/swiftsc.html}, showing that a significant fraction of the X-ray afterglows of GRBs exhibit a shallow decay phase lasting $10^{2}$--$10^{4}$\hspace{0.5mm}s \citep{Zhang_06,Fan:2006zx,Nousek_06,Liang:2007ti,Yu:2007ph,Yamazaki_09}. A number of studies have suggested that the long duration afterglow plateau may be powered by a newly born millisecond magnetar. This could channel slowly decreasing rotational energy into a relativistic outflow via magnetic dipole emission \citep{Usov:1992zd,ZhangNeszaros_2001,Fan:2006zx,DallOsso_2010, Lyons_2010,yu_2010}.

From the perspective of GW detection, the relative local proximity of observed SL-GRBs makes these sources appealing, and raises the possibility of multi-messenger observations by second and third generation GW detectors \citep{KochanekPiran_93,Meszaros_03,abbott_08,Andersson_09,Bloom_2009,Corsi:2009,Corsi:2009b,LSC_SGRBs_2010,Abbott_S5Bursts}. Any detection scenario, of course, depends on the frequency of events in the nearby Universe. Radio observations of 68 local SNe Ib/c by \citet{Soderberg_06a} show that less than 10\% were associated with LGRBs. Based on the local rate of Type Ib/c supernovae  \citep{sodoburg_06_LLGRBRate_06,guettaDellaValle_07} this yields an extreme upper limit of $\sim 2 \times 10^{3}\,\mathrm{Gpc}^{-3}\mathrm{yr}^{-1}$ for SL-GRBs. Studies of SL-GRB rates over the last four years (to be discussed in the next section) have yielded estimates extending over a range $(40-1800)\hspace{1mm} \mathrm{Gpc}^{-3}\mathrm{yr}^{-1}$. These estimates are orders of magnitude greater than those of classical LGRBs.

Although these rate estimates are encouraging, GW detections will also depend on the strength of the emissions. As LGRBs require rapid rotation to produce an accretion disc \citep{YoonLanger_05,WoosleyJankaNature_05,Janka_07,Hartmann_2010}, it is logical to consider post-collapse GW emission mechanisms equally dependent on rotation.

A first possibility is that the strong magnetic fields of newly born magnetars, of order $10^{14}-10^{16}$ G \citep{DuncanThomson_92}, could lead to deformations that would dominate any flattening due to a fast rotation \citep{OstrikerGunn_69,Palomba_00,Konno:2000}. If the deformation axis is offset from the spin axis, this could lead to GW emissions \citep{Palomba_01,Cutler_MagMountains_02,stella_05aa,Regimbau:2006,DallOsso_07,DallOsso_09}.

A second emission possibility, suggested by \citet{Corsi:2009,Corsi:2009b}, is that GW emissions could accompany the electromagnetic dipole emissions of a newly formed magnetar via a secular bar mode instability \citep{Chandrasekhar_70,FriedmanSchutz_CFS_78,LaiShapiro_1995,Ou_barmode_04,shibataBarmode04}.  As this instability occurs on a long timescale $\sim 10^{2}$--\,$10^{4}$\hspace{0.5mm}s, it corresponds well with the observed X-ray plateau of some LGRBs. We note that we do not consider the GW emissions from R modes \citep{owen_r-modes_98,Ferrari_NS_BG_98}, as the effect of the magnetic field could suppress this GW instability in magnetars \citep{Rezzolla_01,Mendell_01}.

Despite the observed sample of SL-GRBs being small, and the origin of this class of bursts still some way from being clearly understood, the rate estimates based on current observations are of an equivalent order to those of other sources of potentially detectable GW backgrounds, e.g. NS/NS mergers \citep{regimbau_NSBG_ApJ_06,regimbau_GWBG_InspNS_07,regimbau_astBG_08}. For this reason we are motivated in this paper to determine if a GW background from SL-GRBs, based on the two mechanisms outlined above, could produce a detectable signal for advanced GW interferometric detectors such as ALIGO\footnote{LIGO---http://www.ligo.caltech.edu/} (Advanced Laser Interferometer Gravitational-wave Observatory) and VIRGO\footnote{VIRGO--- http://www.virgo.infn.it/} or third generation instruments such as the Einstein Telescope\footnote{ET---http://www.et-gw.eu/} (ET). Although we expect there to be variation in the single-source emission mechanisms, we take these models to represent average values. This assumption is reasonable, based on the fact that detection of a stochastic GW background can only yield information on the mean event emission of a population \citep{regimbau_astBG_08}. Additionally, we allow for uncertainties in both the event rates of SL-GRBs and in the frequency of occurrence of the two GW emission mechanisms considered by employing widely separated upper and lower limits.

The possibility of a detectable continuous astrophysical background signal is important, as it could mask the relic GW background signal from the earliest epochs of the Universe. Primordial backgrounds are expected to be produced by large numbers of dynamical events in the early Universe \citep{Grishchuk:1974ny} -- this signal is expected to be isotropic, stationary and unpolarized. An upper limit on the energy density of the primordial GW background normalized by the critical
energy density of the universe, was recently set as $<6.9 \times 10^{-6}$ in the frequency band $(41.5-169.25) \hspace{1mm}\rm{Hz}$ by the LIGO Scientific Collaboration and Virgo Collaboration using data from the S5 two-year science run \citep{LIGO_limit}. This limit improved on previous indirect limits from the big bang nucleosynthesis and cosmic microwave background at around 100 Hz. As ET will be able to detect GW background signals around six orders of magnitude below this limit, it is possible that astrophysical GW background signals could form an additional `noise' component concealing the background signal from primordial processes.

The organization of the paper is as follows: In Section 2 we discuss the rate estimates of SL-GRBs and in Sections 3 and 4, we describe in more detail the two previously mentioned GW emission mechanisms that could result from newly formed magnetars in SL-GRBs. In Section 5 we describe how we will calculate a GW background spectrum and in Section 6 we discuss issues relevant to detection. In Sections 7 and 8 we present our estimations of the GW background signal from our two single-source emission mechanisms and finally draw our conclusions in Section 9.

\section{Rate estimates of sub-luminous GRBs}

\begin{table*}
\label{table_rates}
 \centering
 \begin{minipage}{140mm}
  \caption{A sample of the published estimates on the rate of SL-GRBs along with a brief description of how the rate was determined. }
  \begin{tabular}{@{}lll@{}}
  \hline
    \hline
Reference    &Rate estimate         &\hspace{3cm}Notes               \\
&$\mathrm{Gpc}^{-3}\mathrm{yr}^{-1}$ & \\
\hline
\citet{sodoburg_06_LLGRBRate_06}       &  $230^{+490}_{-190}$
&   A Poisson statistical estimate based on the detection volumes for\\
&&  bursts similar to GRB 980425 and XRF 060218.\\


\hline
\citet{Pian_LLGRBs_06}       &  $110^{+180}_{-20}$
& A fit to the log \emph{N} -- log \emph{P} distribution of BATSE\hspace{0.5mm}\footnote{BATSE - the Burst and Transient Source Experiment on the Compton Gamma-Ray Observatory, launched in 1991, recorded 2704 GRBs during its 9 years of operation.} data using\\
&& a smoothed broken power law LF with a lower bound set by GRB 980425.\\


\hline
\citet{guettaDellaValle_07}       &  $380^{+620}_{-225}$\hspace{0.5mm}\footnote{Estimate based on Poisson statistics.}
& Poisson statistical estimate determined as well as two fits to the \\
&200-–1800\hspace{0.5mm}\footnote{Estimate based on BATSE data.}& log \emph{N} -- log \emph{P} distributions of both \emph{Swift} and  BATSE. A single power law\\
&110-–1200\hspace{0.5mm}\footnote{Estimate based on \emph{Swift} data.}&  LF was used with a lower bound based on GRB 980425.\\

\hline

\citet{Liang_07}       &  $325^{+352}_{-177}$
& The LF and rate density are estimated using \emph{Swift} bursts with known $z$. \\
\hline

\citet{Chapman_07}       &  $700^{+360}_{-360}$
& Estimates obtained by correlating galaxies within 155 Mpc to BATSE\\
&& bursts with properties similar to known SL-GRBs. \\
\hline


\citet{Virgilii_LLGRBs_08}       &  $200^{+200}_{-100}$
& LF parameters and $z$ values estimated through simulation. Rate estimates\\
&& obtained through statistical comparison with the observed \emph{Swift}\\
&&  luminosity--$z$ distribution.\\

  \hline
\hline
\end{tabular}
\end{minipage}
\end{table*}

Table \ref{table_rates} shows rate estimates of SL-GRBs from studies spanning the past four years. Estimates are generally determined by statistical arguments, or fits to the log \emph{N} -- log \emph{P}, peak flux, or `brightness distribution' of bursts. Statistical arguments are typically based on the two closest sub-luminous bursts: GRB 980425 and GRB 060218, detected within 2 years of operation by \emph{Swift}. As rates based on Poisson statistics could be affected by small number statistics, some authors choose to fit to the log \emph{N} -- log \emph{P} distribution of observed bursts. Using this method, a SL-GRB population can be accounted for by decreasing the lower bound of the luminosity function (LF) or by employing a two-component LF \citep{coward_LLGRBs_05}.

The table shows that estimates extend over a range $(40-1800)\hspace{1mm} \mathrm{Gpc}^{-3}\mathrm{yr}^{-1}$, reflecting the present uncertainties on the nature of these bursts. For example, other than uncertainties in the LF, it is still not clear if
these bursts are LGRBs viewed off-axially or are an intrinsically different population \citep[see discussion in][]{coward_LLGRBs_05}. Assuming that SL-GRBs are a unique population with an intrinsic difference in central engine from LGRBs, the rates we will adopt in this study are shown in Table \ref{table_rates}.

For a plausible rate we take the most recent estimate of \citet{Virgilii_LLGRBs_08},
$r_{\rm{P}}= 200 \hspace{1mm} \mathrm{Gpc}^{-3}\mathrm{yr}^{-1}$. As shown in Table \ref{table_rates}, this estimate is of a similar order to the most likely values published in the other studies. As an upper limit we take the largest estimate shown in Table \ref{table_rates} of $r_{\rm{U}}= 1800\hspace{1mm} \mathrm{Gpc}^{-3}\mathrm{yr}^{-1}$ \citep{guettaDellaValle_07} -- this value is $\sim 9\%$ of the local rate of SNe Ib/c. We note that this is a similar fraction of SNe Ib/c producing magnetars to that of Type II SNe -- around 10\% -- as suggested by \citet{murase_06} and \citet{sodoburg_06_LLGRBRate_06}. For our lower bound we take a value of $r_{\rm{L}} = 40 \hspace{1mm} \mathrm{Gpc}^{-3}\mathrm{yr}^{-1}$. We obtain this value by taking a typical lower bound of around $100 \hspace{1mm} \mathrm{Gpc}^{-3}\mathrm{yr}^{-1}$ based on the estimates shown in Table \ref{table_rates} and in correspondence with LGRBs observed during the \emph{Swift} Era, we assume that 40\% of SL-GRBs will also have X-ray plateaus \citep{Evans_09}.

\begin{table}
\centering
\begin{tabular}{lc}
  \hline
 & Rate estimate in \\
  & $\mathrm{Gpc}^{-3}\mathrm{yr}^{-1}$ \\
  \hline
  $r_{\rm{U}}$ & 1800\hspace{1mm} \\
  $r_{\rm{P}}$ & 200\hspace{1mm} \\
  $r_{\rm{L}}$ & 40\hspace{1mm} \\
  \hline
\end{tabular}
   \caption{The sub-luminous GRB rate estimates used in this study. The estimates are denoted by upper, $r_{\rm{U}}$, plausible, $r_{\rm{P}}$, and lower, $r_{\rm{L}}$. }
 \end{table}

\section{Triaxial GW emission from magnetars}
\label{sec_GWmagnetars}
Rotating NSs with a triaxial shape have a time varying
quadrupole moment and hence radiate GWs at a frequency, $f$, which is twice the rotational
frequency.
A  NS born with a rotational period $P_0$ looses rotational energy through magnetic dipole torques and GW emission:
\begin{equation}
\frac{\mathrm{d}E_{\mathrm{rot}}}{\mathrm{d}t} = \frac{\mathrm{d}E_{\mathrm{dip}}}{\mathrm{d}t} +\frac{\mathrm{d}E_{\mathrm{gw}}}{\mathrm{d}t}\,,
\end{equation}
with rotational, dipole and gravitational energy loss rates:


\begin{equation}
\label{eq-energy}
\left\{\begin{array}{l}
\displaystyle \frac{\mathrm{d}E_{\mathrm{rot}}}{\mathrm{d}t} = \displaystyle \pi^2 I_{\mathrm{zz}} f \frac{\mathrm{d}f}{\mathrm{d}t}, \\\\
\displaystyle\frac{\mathrm{d}E_{\mathrm{dip}}}{\mathrm{d}t} = \displaystyle K_{\mathrm{dip}} f^4  =\frac{\pi^4 R^6B^2}{6c^3}f^4\,,\\\\
\displaystyle\frac{\mathrm{d}E_{\mathrm{gw}}}{\mathrm{d}t} = \displaystyle K_{\mathrm{gw}} f^6=\frac{32\pi^6 GI_{\mathrm{zz}}^2 \rho^2}{5c^5}f^{6}\,.\\
\end{array}
 \right.
\end{equation}
These result in a change in the frequency:
\begin{equation}
\frac{\mathrm{d}f}{\mathrm{d}t} = \frac {K_{\mathrm{dip}}}{\pi^2 I_{\mathrm{zz}}} f^3 + \frac {K_{\mathrm{gw}}}{\pi^2 I_{\mathrm{zz}}} f^5
\label{eq-fdot}
\end{equation}
From the equations for $\mathrm{d}E_{\mathrm{gw}}/\mathrm{d} f$ and $\mathrm{d}f/\mathrm{d}t$, we can write the emitted GW spectral energy density as follows:
\begin{equation}
\frac{\mathrm{d}E_{\mathrm{gw}}}{\mathrm{d} f} = K f^3 (1+\frac{K}{\pi^2 I_{\mathrm{zz}}} f^2)^{-1} \,\ \rm{with}\,\ \emph{f} \in \lbrack 0, 2/\emph{P}_0 \rbrack
\label{eq-Enj_pulsar}
\end{equation}
where
\begin{equation}
K = \frac{192 \pi^4 G I_{\mathrm{zz}}^3}{5 c^2 R^6} \frac{\rho^2}{B^2}\,.
\label{eq-K_pulsar}
\end{equation}

\noindent In this expression $R$ is the radius of the star, $\rho=(I_{\mathrm{xx}}-I_{\mathrm{yy}})/I_{\mathrm{zz}}$ is the ellipticity in terms of the principal moments of inertia, $B = B_s \sin \alpha$ where $B_s$ is the surface polar magnetic field strength and $\alpha$ the angle between the rotational and dipole axes.
%

The majority of NSs are understood to be born with magnetic fields of the order of $10^{12}-10^{13}$ G and rotational periods of the order of tens or hundreds of milliseconds \citep{Regimbau:2000,Faucher:2006,Soria:2008} and will make a negligible contribution. However, NSs with sufficiently high initial rotational periods, $\sim$ 1--3\hspace{0.5mm}ms, which undergo violent convection or differential rotation during the first seconds after birth can obtain super-strong crustal magnetic fields  ($B_s$ in the range $\sim 10^{14}-10^{16}$ G) through an $\alpha$--$\Omega$ dynamo action  \citep{DuncanThomson_92,Thompson:1993hn}.
For these highly magnetized NSs, the distortion induced by the magnetic torque can
become significant \citep{Palomba_00, Konno:2000}, and GW emission can be orders of magnitudes larger than for ordinary NSs \citep{Palomba_01}. Because it could carry away most of the initial rotational energy of millisecond magnetars, this scenario provides a natural explanation for the absence of the signature of enhanced energy injections in X-ray spectra of supernova remnants around known magnetars \citep{DallOsso_09}.

In this study, we  take $P_0=1$ ms, and corresponding to a typical NS of mass $1.4M_{\odot}$ take $R = 10$ km and  $I_{\mathrm{zz}}=1 \times 10^{38}$ kg m$^2$ \citep{ArnettBowers76,Bonazzola:1995}. We consider two different scenarios corresponding to different configurations of the magnetic field:
\begin{enumerate}
\item \emph{Poloidal field}
For the case in which the internal magnetic field is purely poloidal and matches the dipolar field in the exterior, the ellipticity is given by \citet{Bonazzola:1995,Konno:2000}:
\begin{equation}
\rho_B=g \frac{R^8B^2}{4GI_{\mathrm{zz}}^2}\,.
\label{eq-epsB_pol}
\end{equation}
According to the numerical simulations of \cite{Bonazzola:1995}, the magnetic distortion
parameter $g$ of a typical NS, which depends on both the equation of state and
the magnetic field geometry, can range from $1-10$ for a
non-superconducting interior and can increase to $100-1000$ for a type I superconductor in which all the magnetic field has been expelled from the superconducting core. It can take on an even greater values for type II superconducting cores or counter rotating electric currents.
Following \cite{Regimbau:2006} we take $ g =520$ in the mid-range of permissible values of a type I superconducting core, corresponding to a core whose dimension represents $95\%$ of the equator radius. We take $B = 5 \times 10^{14}$ G as representative of the magnetar population, based on average observational values of soft gamma repeaters and anomalous X-ray pulsars\footnote{The McGill SGR/AXP online catalog can be found at
http://www.physics.mcgill.ca/~pulsar/magnetar/main.html}, but note that we have excluded two objects with characteristic times of 230 kyr and $>1300$ kyr for which dissipation of the magnetic field may have been important.

We find ellipticity $\rho_B = 4.8 \times 10^{-4}$. To have an idea of the magnitude of such deformation, we can compare this ellipticity with the ones predicted for elastic deformations of NSs. In this respect, $\rho_B = 4.8 \times 10^{-4}$ is about two orders of magnitude larger than the maximum elastic quadrupole deformation of conventional NSs \citep{Horowitz2010}, but comparable to elastic deformations sustainable by solid strange stars, and $1-2$ orders of magnitude below the upper limit derived for crystalline color-superconducting quark matter \citep{Owen_05,Lin2007}.
For $\rho_B = 4.8 \times 10^{-4}$, the GW emission is negligible compared to the magnetic torque and eq.~(\ref{eq-Enj_pulsar}) simplifies to ($K<<\pi^2 I_{\mathrm{zz}} f^{-2}$):
\begin{equation}
\frac{\mathrm{d}E_{\mathrm{gw}}}{\mathrm{d} f} \sim K f^3\,.
\label{eq-Enj_pulsar2}
\end{equation}
We note that for a rotation period of the order of ms, strongly magnetized relativistic winds could slow down the star in a few minutes as energy is rapidly transferred to the ejecta \citep{tho04,buc06,met07}. However, this effect is expected to be negligible for  $B<(6-7) \times 10^{14}$G.

In theory, one could consider values up to $g = 1000 - 10000$ and $B = 10^{16}$ G for which GW emission would dominate in most of our frequency range ($K >> \pi^2 I_{\mathrm{zz}} f^{-2}$). However, this scenario would produce ellipticities of order unity, much higher than the ones typically considered for magnetic \citep{DallOsso_09} or elastic \citep{Owen_05} NS deformations.

\item \emph{Toroidal field}
A number of studies have suggested that the internal magnetic field $B_t$ could be dominated by a strong toroidal component in the range $10^{15}-10^{17}$ G \citep{Cutler_MagMountains_02,stella_05aa,DallOsso_07,DallOsso_09}, which could induce a prolate deformation with ellipticity
\begin{equation}
\rho_B \sim 1.6 \times 10^{-4} \left(\frac {B_{t}}{10^{16}}\right)^2\,.
\label{eq_B_tor}
\end{equation}
Following \cite{DallOsso_09}, we assume a pure internal toroidal field of $B_t = 10^{16}$G , with an external magnetic field of the order $10^{14}$ G.
$B_t$ was deduced by \cite{stella_05aa} from studies of the energetics and likely recurrence time of the 2004 December 27 event from SGR 1806-20, and is consistent with the thermal emissions observed in Anomalous X-ray Pulsars, assuming they are powered by the decay of the magnetic field \citep{kam07}. This value also supports constraints set by \cite{vink_Kuiper_2006} on the X-ray spectra of the supernova remnants surrounding known magnetars.

In this case, both GW and magnetic dipole losses contribute to the magnetar spindown. At small frequencies ($f \lesssim 100$\hspace{0.5mm}Hz), the emission is dominated by the electromagnetic contribution, but above $f \sim 500$\hspace{0.5mm}Hz, GW emission becomes the most important process, approaching the saturation regime where spindown is purely gravitational ($K>>\pi^2 I_{\mathrm{zz}} f^{-2}$):
\begin{equation}
\frac{\mathrm{d}E_{\mathrm{gw}}}{\mathrm{d} f} \sim \pi^{2} I_{\mathrm{zz}} f\,.
\label{eq-Enj_pulsar3}
\end{equation}
Increasing $B_t$ to $5 \times 10^{16}$G or $10^{17}$G lowers the	frequency at which GW emission becomes the dominant contribution to around 100\hspace{0.5mm}Hz or 30\hspace{0.5mm}Hz.
\end{enumerate}

\section{GW emission from secular bar mode instabilities}
\label{BarModeInst}

Bar-mode instabilities associated with NS formation derive their name from the `bar-like' deformation they induce, transforming a spheroidal body into an elongated bar that tumbles end-over-end. The highly non-axisymmetric structure resulting from a compact astrophysical object undergoing this instability makes such an
object a potentially strong source of gravitational radiation \citep{Chandrasekhar_69,Chandrasekhar_70,FriedmanSchutz_CFS_78,LaiShapiro_1995,brown00,nct00,sbs00,Ou_barmode_04,shibataBarmode04,Baiotti_barmode_PRD_07,Dimmelmeier_SNCC_08,ott_review_09}.

A system susceptible to bar-mode deformation is parametrized by the stability parameter, $\beta = T/|W|$, where $T$ is the rotational kinetic energy and $W$ is the gravitational potential energy.

There exist two different
timescales and mechanisms for these instabilities. Uniformly rotating, incompressible stars are \emph{secularly} unstable if $\beta \gtrsim 0.14$, and have a growth time that is determined by the time-scale of dissipative processes in the system (such as viscosity or gravitational radiation) -- usually much longer than the dynamical time-scale of the system \citep[see e.g.][]{ssbs01, nct00}. In contrast, a \emph{dynamical} instability sets in when $\beta \gtrsim 0.27$, and has a growth time of the order of the rotation period of the object \citep[see e.g.][]{nct00}. This is expected to be the fastest growing mode.

Because bar-mode instabilities are a potentially important source of gravitational radiation, they have been the subject of many numerical studies. \cite{Dimmelmeier_SNCC_08} found that the post-bounce core cannot reach sufficiently rapid rotation to become unstable to the classical high-$\beta$ dynamical bar-mode instability. However, they found that many of their post-bounce core models had sufficiently rapid rotation to
become subject to a low-$\beta$ dynamical instability first by seen by \citet{Centrella_lowBeta_01}. The potential for enhancements in the GW emissions by dynamical instabilities at low $\beta$ is encouraging and has been demonstrated in a number of other studies \citep{ske02,ott_lowbeta_05,Scheidegger_lowbeta_08,Scheidegger_lowbeta_2010}.

The requirement of rapid rotation for post-collapse instabilities suggest that a GRB progenitor, typically required to be in high rotation \citep{WoosleyJankaNature_05}, may provide favourable conditions for such instabilities to set in. In this paper we follow \citet{Corsi:2009,Corsi:2009b} and consider GW emissions from the longer lived secular bar-mode instability possibly associated with the observed shallow decay phase observed in GRB X-ray afterglows discussed earlier. In the next section we describe the single source models we employ to estimate the GW backgrounds from this instability.

\subsection{Single-source spectrum from secular bar-mode instabilities}

The single-source emission mechanism used in this section is motivated by the study of \citet[][]{Corsi:2009} who estimated the GW emissions from a secular instability in a newly born magnetar. Their work extended the work of \citet{LaiShapiro_1995} for the quasi-static evolution of NSs under gravitational radiation. Treating the NS as a polytrope of index $n=1$, they assumed typical parameter values: total mass, $M = 1.4 M_\odot$; radius, $R= 20\hspace{0.5mm}\rm{km}$; initial magnetic dipole field strength at the poles, $B = 10^{14} G$; and $\beta = 0.20$ corresponding with the middle of the expected range for the secular instability ($0.14 < \beta < 0.27$). They estimated quasi-periodic GW emissions at around 150\,Hz, with characteristic amplitudes $h_c \sim 10^{-21}$ at 10\hspace{0.5mm} Mpc.

\begin{figure}
\begin{center}
\includegraphics*[bbllx = 60pt,bblly =260pt, bburx = 550pt, bbury =590pt,scale = 0.50]{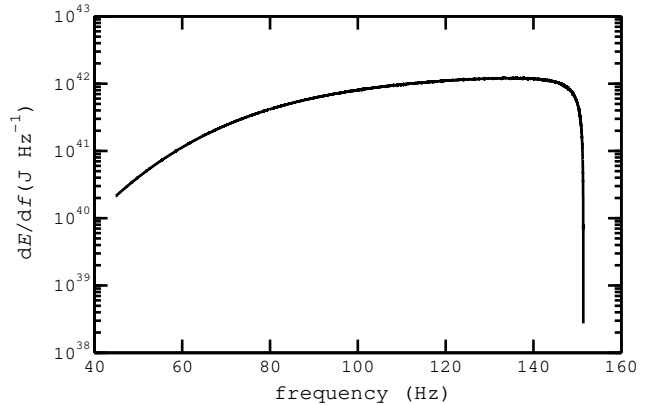}
\end{center}
\caption[]{The rest frame energy spectrum for GW emission by a secular bar mode instability for parameter values $M = 1.4 M_\odot$; $R= 20\hspace{0.5mm}\rm{km}$; $B = 10^{14}\hspace{0.5mm}\rm{G}$; $\beta = 0.20$. During the first 2000\hspace{0.5mm}s of the evolution, the GW energy is emitted in the 50--150 Hz energy band.} \label{fig_dedf_secular}
\end{figure}

Figure \ref{fig_dedf_secular} shows the rest frame energy spectrum, $\mathrm{d}E_{\mathrm{gw}}/\mathrm{d}f$, from this mechanism. This function was computed through

\begin{equation}\label{eq_dedf_secular}
    \frac{\mathrm{d}E_{\mathrm{gw}}}{\mathrm{d}f} = \frac{\mathrm{d}E_{\mathrm{gw}}}{\mathrm{d}t}\large\left|\frac{ \mathrm{d}t}{\mathrm{d}f} \right|\,,
\end{equation}

\noindent using data for the luminosity and frequency evolution of the instability supplied by \citet{Corsi:2009}. The bulk of the emission takes place over the first 2000\,s; during the first 100\,s the GW frequency is constant at 150 Hz. This is then followed by a slow decline over about 2000\,s to $\sim 100$ Hz. Their analytical treatment included the effects of energy losses from magnetic dipole radiation. In principle, the secular evolution would bring the star to reach a stationary football configuration; thus one could follow the predicted GW signal until its frequency approaches zero. \citet{Corsi:2009} conservatively shut off the bar emission after about few thousand seconds of evolution (as for typical GRB plateau durations), when the GW signal had a frequency of about 50\hspace{0.5mm}Hz, and its amplitude was falling below the ALIGO sensitivity curve. We note that the ET sensitivity below 50\hspace{0.5mm}Hz is one order of magnitude better than for ALIGO, but the final stages of the secular evolution are indeed highly uncertain (due to e.g. viscosity effects possibly coming into play, see e.g. Lai and Shapiro 1995, Corsi and Meszaros 2009). Thus, also in this analysis we conservatively assume that the bar emission completely shuts off on the typical duration of GRB plateaus, when the GW signal frequency is around 50 Hz.

We note here that, despite uncertainties on the values of the model
parameters, Fig.\ref{fig_dedf_secular} can be considered as an average case: mass, radius and
magnetic field of the NS are chosen so as to represent the typical case
for a newborn magnetar; $\beta$, is chosen to
be in the middle of the secular instability range. We also
stress that, as discussed in Corsi \& Meszaros 2009, with this typical
choice of parameters, the observed timescale of GRB plateaus is correctly
reproduced. Moreover, the amount of energy released by the magnetar for
dipole losses during such timescale, is of the order of $10^{50}$ ergs, i.e.
comparable to the isotropic energy output of long sub-luminous GRBs, and
thus sufficient to actually cause a visible plateau in their light
curve.

\section{The Gravitational Wave Background}

\subsection{Cosmic metallicity evolution}

The spectral form of an astrophysical GW background (AGB) is highly dependent on the variation of the event rate with $z$ \citep{howell_GWBG_04,coward_GWBG_01,regimbau_astBG_08}. In general, due to the relative short lifetimes of massive stars (of order tens of Myr) the transient populations that produce AGBs are assumed to track the star formation rate (SFR); for transient populations of coalescing compact objects an additional factor is included to account for merger time delay. For the case of an AGB from SL-GRBs, there is growing evidence that cosmic metallicity evolution must also be considered \citep{Li_Metallicity_08,Modjak_Metallicity_08}.

For WR stars to retain sufficient rotation to power a GRB, angular momentum losses through stellar-wind induced mass-loss must be minimized \citep{Woosley_GRB_Progenitors_06}. As wind-driven mass loss in WR stars is understood to be dependent on a high enough fraction of iron, a low-metallicity environment is an essential requirement \citep{vink_WR_05, WoosleyJankaNature_05}.

To account for metallicity evolution of SL-GRBs with redshift we adopt the simple model of \citet{langerNorman_metalisity_06}

\begin{equation}\label{eq_metal}
    \Psi(z,\epsilon)=\frac{  \hat{\Gamma}(0.84, \epsilon^{2} 10^{0.3 z}) } {\Gamma(0.84)}\,,
\end{equation}

\noindent where $\epsilon = Z/Z_{\odot}$ is the fraction of solar metallicity and $\Psi(z,\epsilon)$ is the fraction of massive stars at $z$ born with metallicity less than $Z_{\odot}\epsilon$. Here, $\hat{\Gamma}$ and $\Gamma$  are the incomplete and complete gamma functions. \citet{langerNorman_metalisity_06} found that a metallicity cutoff of $\epsilon = 0.1$, corresponding to $\sim 1$ GRB per 100 WR stars throughout the Universe, was able to reproduce the observed global ratio of the rates of GRBs to core-collapse SNe. This value was also used by \citet{salvaterra_07} who suggested that luminosity evolution was required to reproduce the \emph{Swift} distribution at high $z$. An analysis of the \emph{Swift} data (to August 2009) by \citet{Butler_2010} ruled out luminosity evolution and found a more relaxed cutoff $\epsilon \sim 0.2 - 0.5$, was adequate to reproduce the observed sample. Their more modest dependence on metallicity was supported by studies of the mass distribution of GRB host galaxies by \citet{Kocevski_09}. Based on these studies, we adopt here a range $\epsilon \sim 0.1 - 0.5$ to allow for present uncertainty.

\subsection{Source rate density evolution}

Our source rate evolution model for SL-GRBs with redshift, $R_{\rm SL}(z)$, is obtained by scaling the star formation history\footnote{In units of mass converted to stars per unit time and volume.}, $R_{\rm SF}(z)$, with the function $\Psi(z, \epsilon)$:

\begin{equation}\label{eq_sfr}
    R_{\rm SL}(z)= \Psi(z, \epsilon) R_{\rm SF}(z)\,.
\end{equation}

\noindent For $R_{\rm SF}(z)$, we employ the model of \cite{hb_sfr_06}, who constrained the star formation rate history by combining recent measurements taken from sources observed at ultraviolet, far-infrared and radio with previous more robust measurements taken over the last decade. By normalizing $R_{\rm SL}(z)$ to the the local $(z = 0)$ rate, we produce a dimensionless evolution factor

 \begin{equation}\label{eq_ez}
    e(z)= R_{\rm SL}(z)/R_{\rm SL}(z=0)\,.
 \end{equation}

 \noindent This allows us to extrapolate a local rate density to cosmological volumes.

\begin{figure}
\begin{center}
\includegraphics[ scale = 0.55]{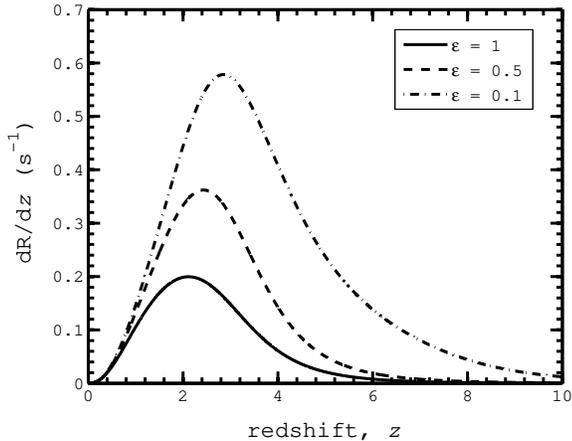}
\end{center}
\caption[]{The differential event rate $\mathrm{d}R/\mathrm{d}z$ of SL-GRBs as a function of redshift based on the SFR
 model of \citet{hb_sfr_06}. The three curves correspond to different metallicity cutoffs $\epsilon = (0.1;0.5;1) Z_{\odot}$ at the upper rate estimate of $r_{\rm{U}}= 1800\hspace{1mm} \mathrm{Gpc}^{-3}\mathrm{yr}^{-1}$. Inclusion of cosmic metallicity evolution increases the contribution from sources at higher $z$. This is shown by a shift in the peak of $\mathrm{d}R/\mathrm{d}z$ from $z \sim 2$ ($\epsilon = 1$) to $z \sim 3$ assuming SL-GRBs follow a low metallicity dependence of $\epsilon = 0.1 Z_{\odot}$.} \label{fig_drdz}
\end{figure}

\subsection{The event rate equation}

In order to evaluate the contribution by a population of GW sources to a stochastic background signal, knowledge of the rate of emissions from recent and past epochs is essential. The source rate evolution of an AGB is modeled by the differential event rate, $\mathrm{d}R /\mathrm{d}z$, which describes the rate of events in the redshift shell $z$ to $z+{\mathrm d}z$:

\begin{equation}\label{drdz}
\mathrm{d}R = \frac{\mathrm{d}V}{\mathrm{d}z}\frac{r_0 e(z)}{1+z} \mathrm{d}z \,.
\label{eq_drdz}
\end{equation}

\noindent The $(1 + z)$ factor accounts for the time dilation of the observed rate by cosmic expansion; its inclusion converts source-count information to an event rate. The parameter $r_{0}$ is the local rate density, usually defined within a volume spanning the Virgo cluster of galaxies (at around 20 Mpc). This factor is fundamental to estimating the number of potentially observable events and is determined using estimated source rates within a larger fixed volume of space. For the factor $r_{0}$, we adopt the values discussed in section 2 of $(r_{\rm{U}},r_{\rm{P}},r_{\rm{L}})=(1800,200,40)\hspace{1mm} \mathrm{Gpc}^{-3}\mathrm{yr}^{-1}$.

The co-moving volume element $\mathrm{d}V$ describes how the number densities of non-evolving objects locked into Hubble flow are constant with redshift. This is obtained by calculating the luminosity distance from \citep[cf.][p.\hspace{0.5mm}332]{Peebles}

\begin{equation}\label{lumd}
d_\mathrm{L}(z) = (1 + z)\frac{c}{H_{0}}\int_{0}^{z} \frac{\mathrm{d}z\hspace{0.5mm}'}{h(z\hspace{0.5mm}')}\,,
\end{equation}

\noindent and using eq.\hspace{1.0mm}3 of \citet[][]{pm01},

\begin{equation}\label{dvdz}
\frac{\mathrm{d}V}{\mathrm{d}z}= \frac{4\pi c}{H_{0}}\frac{d_\mathrm{\hspace{0.25mm}L}^{\hspace{1.5mm}2}(z)}{(1 +
z)^{\hspace{0.25mm}2}\hspace{0.5mm}h(z)}\hspace{2.0mm}.
\end{equation}

\noindent The normalized Hubble parameter, $h(z)$, is given by

\begin{equation}\label{hz}
h(z)\equiv H(z)/H_0 = \big[\Omega_{\mathrm m} (1+z)^3+ \Omega_{\mathrm \Lambda} \big]^{1/2}\,,
\end{equation}

\noindent for a `flat-$\Lambda$' cosmology. We take $\Omega_{\mathrm
m}=0.3$, $\Omega_{\mathrm \Lambda}=0.7$ and
\mbox{$H_{0}=70$ km s$^{-1}$ Mpc$^{-1}$} for the Hubble parameter at the present epoch \citep{Rao_737Cos_06}.

Figure \ref{fig_drdz} shows the form of d$R$/d$z$ assuming different metallicity cutoffs for SF-GRBs. In this plot we compare curves corresponding to the different cutoffs $\epsilon = (0.1,0.5)$ with the $\epsilon = 1$ solar metallicity curve. The inclusion of cosmic metallicity evolution increases the contribution from higher-$z$ sources, shown by a shift of the peak from $z \sim 2$ to $z \sim 3$ for a low metallicity dependence, $\epsilon=0.1Z_{\odot}$. The magnitude of the peak value also increases by up to a factor of 3. Thus, the additional contribution from sources at higher $z$ should enhance the lower frequency component of the stochastic background signal through redshift.


\subsection{The Gravitational Wave background spectra}

The spectral time-integrated flux density or spectral fluence, in J m$^{-2}$ Hz$^{-1}$, of a quadrupole GW
signal at a luminosity distance $d_{\mathrm{L}}(z)$ from a single source can be expressed as

\begin{equation}
 F_{\mathrm{ss}} (f_{\rm{obs}},z) = \frac{\mathrm{d}E_{\mathrm{gw}}(f_{\rm{obs}})}{\mathrm{d}f}\,\frac{(1 + z)}{4 \pi d_{\mathrm{L}}(z)^{2}}\,,\label{flux}
\end{equation}

\noindent where $\mathrm{d}E_{\mathrm{gw}}(f_{\rm{obs}})$ is the spectral GW energy at the observed frequency $f_{\rm{obs}}$, which is related to the source frequency $f$ by the
redshift factor: \mbox{$f_{\rm{obs}} = f \mathord{\left/ {\vphantom {f_{\rm{obs}} {(1 + z)}}} \right.
\kern-\nulldelimiterspace} {(1 + z)}$.}

The background spectral flux density, in W m$^{-2}$ Hz$^{-1}$, from all events throughout the Universe is
obtained by integrating the product $F_{\mathrm{ss}}( f_{\rm{obs}}, z) \mathrm{d}R/\mathrm{d}z$ over the
redshift range $z = 0$ to $z_{\mathrm{lim}}$:

\begin{equation}
  F_{B}(f_{\rm{obs}})=\int_{0}^{z_{\mathrm{lim}}} [\,F_{\mathrm{ss}}(f_{\rm{obs}},z)(\mathrm{d}R/\mathrm{d}z)\,]\hspace{0.1cm}\mathrm{d}z\,,\label{FB}
\end{equation}

\noindent with the integrand given by (\ref{drdz}) and (\ref{flux}) within a limiting redshift, $z_{\mathrm{lim}}=10$, which we take as the beginning of stellar activity. In support of this value we note that GRB 090423, the most distant recorded burst ($z=8.2$), showed no evidence of properties that were inconsistent with the majority of the observed GRB population \citep{Nature_GRBatZ82_2009}.

The spectral energy density of a GW background is conventionally expressed by the dimensionless energy density parameter, $\Omega _{\mathrm{B}}(f_{\rm{obs}})$, defined as the energy density of GWs per logarithmic frequency interval
normalized to the cosmological critical energy density $\rho_c = 3H_{0}^{2}/8 \pi G$. This function can be constructed from $F_{\mathrm{B}}(f_{\rm{obs}})$ \citep{Ferrari_SNBG_98,howell_GWBG_04,regimbau_astBG_08}:

\begin{equation}
\Omega _B (f_{\rm{obs}}) = f\,F_{B}(f_{\rm{obs}}) \mathord{\left/ {\vphantom {{f_{\mathrm{obs}} F_B (f_{\mathrm{obs}} )} (}} \right.
\kern-\nulldelimiterspace} (\rho _c c^3)\,\;. \label{EQns-closure}
\end{equation}

\noindent Throughout this paper we will present our estimated GW background spectra using this function.

\subsection{The duty cycle of an astrophysical GW background}
\label{subsubsection-The Duty Cycle of an Astrophysical GW background}

When we consider AGB signals, in addition to the energy density and characteristic frequency, another useful quantity is the duty cycle, $DC$ \citep{blair_SNBG_96, magg00, coward_regimbau_06,regimbau_astBG_08}. The value of the $DC$ is given by the ratio of the typical duration
of the signal, $\tau$, to the average waiting time between the reception of successive events. The waiting time will depend on the rate of events as observed in our frame, $R$, and thus a $DC$ is generally defined by the quantity $R \times \tau$. When considering a cosmological distribution of events, the $DC$ is determined by sources within a limiting redshift, $z_{\mathrm{lim}}$:

\begin{equation}
DC(z_{\mathrm{lim}}) = \int_{0}^{z_{\mathrm{lim}}}  (1 + z)\hspace{0.2mm}\tau\hspace{0.2mm}(\mathrm{d}R /\mathrm{d}z)\hspace{0.5mm}\mathrm{d}z\;,\label{eq_DC}
\end{equation}

\noindent Here, the signal duration is dilated to \mbox{$(1+z)\tau$} by the cosmic expansion and the quantity $\mathrm{d}R /\mathrm{d}z$ is the cosmologically dependent differential event rate given by equation \ref{drdz}.

Many studies are concerned with the value of \emph{DC} as seen at the detector and determine the value of $DC$ by setting $z_{\mathrm{lim}}$ equal to the redshift at which stellar activity began. In this case equation (\ref{eq_DC}) provides a total value, $DC(z_{\mathrm{lim}})$. As source rate evolution will increase out to large cosmological volumes, it is interesting to see how \emph{DC} too increases with $z$. In this study, we will therefore calculate $DC$ as a function of redshift.

In general, for an AGB the signal is defined as continuous if it has a $DC$ of unity or above. As it will still be possible to resolve individual events in this regime, a more conservative threshold can be defined from the view of single detections as $DC \geq 10$ \citep[see][]{Regimbau:2009rk}. Thus, using the convention of \cite{regimbau_NSBG_ApJ_06}, non-continuous signals can be further categorized into shot and popcorn type. More descriptive definitions of these components are provided as follows:\\

  \textbf{Continuous} ($DC \geq 10$): This signal is, at any given time, the superposition of a large, but random, number of overlapping signals. As the amplitude of each contributing signal is itself random, the central limit theorem will apply leading to a Gaussian distribution in amplitudes. Because it will be difficult to resolve the individual components, this signal can potentially mask a background signal of primordial origin \citep{magg00}. For this reason, AGB with $DC \geq 10$ would be bad news from the perspective of primordial background searches.\\

   \textbf{Popcorn noise} ($0.1\leq DC < 10$): This signal will manifest in GW data as a non-continuous stochastic background signal with an amplitude distribution dependent on the spatial distribution of the sources. For background signals with $DC$s at the lower end of the popcorn noise range, the  individual components will rarely overlap. In this case the signal will be dominated by the closest events of a population of sources.\\

  \textbf{Shot noise} $DC < 0.1$ : For this signal, the waiting time between events is large in comparison with duration of a single event \citep{regimbau_NSBG_ApJ_06}.\\

Using these three definitions of $DC$, \cite{coward_regimbau_06} have shown that an AGB can be divided into three different detection regimes, each defined by the corresponding shells of $z_{\mathrm{lim}}$. For most types of AGB, at low $z$, the signal produced is predominantly of the shot-noise type, extending to popcorn and continuous with $z$, due to time dilation and increased source rate density. The weighting of the different AGB regimes for a particular source population depends strongly on the event rate and is an important consideration when selecting the most appropriate signal detection strategy.

\section{Detection}
\label{section_detection}

In this study we consider the design sensitivities of second generation interferometric detectors, such as ALIGO\footnote{http://iopscience.iop.org/0264-9381/27/8/084006} (expected online in 2015) and Virgo\footnote{https://pub3.ego-gw.it/codifier/includes/showTmpFile.php?doc =2219\&calledFile=VIR-027A-09.pdf} \footnote{www.virgo.infn.it}, and third generation ones, for which we will use ET\footnote{A three year design study for the Einstein GW telescope began in May 2008. For details see http://www.ego-gw.it/ILIAS-GW/FP7-DS/fp7-DS.htm, "Design Study Proposal for E.T. (Einstein Gravitational Wave Telescope)", submitted to the EU Seventh Framework Programme.}(possible construction will be late in the next decade).

\begin{figure}
  \includegraphics[scale = 0.56]{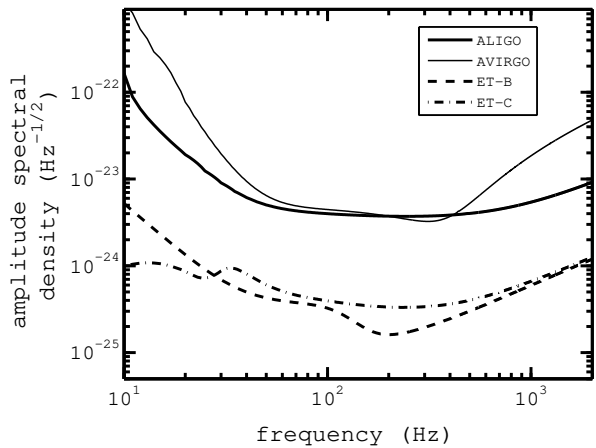}\\
  \caption{The design sensitivity curves for advanced LIGO (ALIGO), Advanced Virgo (AVIRGO) and two possible configurations of the third generation Einstein Telescope (ET): ET-B and ET-C. See Sec. 7 for more details.}\label{fig_noisecurves}
\end{figure}

For ALIGO, we use the sensitivity curve based on the zero detuning, high laser power configuration\footnote{The ALIGO sensitivity curve is described in the public LIGO document
 LIGO-T0900288-v2 (https://dcc.ligo.org/public/0002/T0900288/002/AdvLIGO noise curves.pdf).}. For ET we consider two target sensitivities. Firstly, ET-B, which is an underground based design, incorporating long suspensions, cryogenics and signal and power recycling \citep{HildETconventional}. Secondly, a so-called \emph{Xylophone} configuration, ET-C, which merges the output of two detectors specializing in different frequency bands: a) an underground low-frequency cryogenic configuration with long suspensions and moderate laser power; b) a high frequency detector implementing squeezed light states, large test masses and a high power laser \citep{Hild2010}. The advantage of this strategy is that it can decouple the obstacles in operating a high power laser alongside a cryogenically cooled suspension optimized for thermal noise \citep{Shoemaker2001LIGOXylophone}. The design sensitivity curves for these four detectors are shown in Figure \ref{fig_noisecurves}.

The most promising detection strategy for continuous GW background signals is cross-correlating the output of two
neighbouring detectors \citep[see,][]{magg00,allenRomano}. For this strategy to be achieved, the detectors must be separated by less than one reduced wavelength, which is about 100 km for frequencies around 500 Hz where $\Omega _{\mathrm{B}}(f)$ might peak. The detectors also need to be sufficiently well separated that their noise sources are largely uncorrelated. We note that although this may not be possible for ET, techniques are in development to remove environmental noise and instrumental correlations \citep{Fotopoulos:2008yq}.

Under these conditions, assuming Gaussian noise in each detector and optimal filtering, a filter function chosen to maximize the signal-to-noise ratio, SNR, for two such detectors is given by \citep[][eq.3.75]{allenRomano}

\begin{equation}
\left( \frac{\mathrm{S}}{\mathrm{N}} \right)^{2} \approx \frac{9H_{0}^{4}} {50 \pi^{4}} T \int_{0}^{\infty}
\frac{\gamma^{2}(f)\Omega_{B}^{2}(f)}{f^{6} S_{n1}(f)S_{n2}(f)}\, \mathrm{d}f\,.
\label{snr}
\end{equation}

\noindent Here $\gamma (f)$ is the `overlap reduction function', which accounts for the separation and relative orientation of the detectors, and $S_{n1}(f)$ and $S_{n2}(f)$ are the noise power spectral densities of the detectors; $T$ is the integration time. As the optimal filter depends on $\Omega_{B}(f)$, a range of filter functions based on theoretical expectations of this function will need to be used. In this study we adopt a value of $SNR = 3$ to indicate detection, corresponding with false alarm rate of 10\% and detection rate of 90\% and \citep{allenRomano}.

For signals of the shot noise and popcorn type, the standard cross correlation strategy in the frequency domain may not be optimal and other methods in the time domain have been proposed or are currently under investigation in the LIGO-Virgo collaboration \citep{DrascoFlanagan_NGausGWBG_03,coward_PEH_05}. For signals of the shot noise type, individual events may be clearly distinguishable, and if within a detector's range, can potentially be resolved using data analysis techniques such as matched filtering.

\begin{figure}
\begin{center}
\includegraphics*[scale = 0.6]{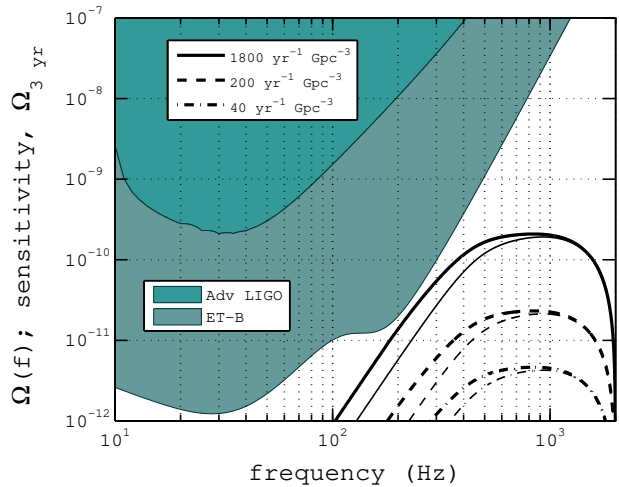}
\end{center}
\caption[]{The GW background spectrum from triaxial deformations in newly born magnetars associated with SL-GRBs in the regime where the spindown is dominated by the magnetic torque. The three curves assume that triaxial deformations are introduced by an internal poloidal magnetic field ($B=5 \times 10^{14}\mathrm{G};g=520;I_{\mathrm{zz}}=10^{45}\mathrm{kg m}^2$;R=10km). The curves are presented for three rates of occurrence: $(r_{\mathrm{U}},r_{\mathrm{P}},r_{\mathrm{L}})= (1800,200,40)\,\mathrm{Gpc}^{-3}\mathrm{yr}^{-1}$ and metallicity cutoffs of $\epsilon = 0.1$ (thick curves) and. $\epsilon = 0.5$ (thin curves). The sensitivity curves of second and third generation laser interferometric detectors are represented by ALIGO and ET-B, in terms of $\Omega_{\mathrm{det}}(f)$ assuming a 3 year integration. Based on observational values of SGRs and anomolous X-ray pulsars, we assume the value of $B$ used here is representative of the magnetar population. This value gives $\rho_{B}=4.8\times10^{-4}$, comparable to elastic deformations sustainable by solid strange stars, and 1-2 orders of magnitude below the upper limit derived for crystalline color-superconducting quark matter. }
\label{fig_magnetar_poloidal}
\end{figure}

To estimate the detectability of the GW backgrounds considered in this paper we will assume continuous signals, and hence use the cross correlation statistic to determine SNRs. For cases in which a significant proportion of the background signal will not be continuous, we will also investigate the different $z$ regimes in which the shot, popcorn and continuous regimes exist. We calculate the SNRs for the two generations of GW detector outlined above; for second generation we assume 3 years of integration by an ALIGO configuration; for third generation we assume 1 year of integration by (ET-B; ET-C). We further assume: a) separated detectors; b) an optimal case in which a pair of equivalent detectors are situated within several km and aligned. For ALIGO we will employ the LIGO Hanford/Livingston pair (H1-L1) for scenario a), using for $\gamma (f)$ the form given by eq. 3.26 of \citet{Allen_2002}. For ET we assume two detectors located in Cascina, of triangular shape ($60^{o}$ between the two arms) and separated by an angle of 120$^{o}$.
In the frequency range we are interested in (1--1000\hspace{0.5mm}Hz), $\gamma (f)$ reduces to a value of --3/8.
For case b) we will assume $\gamma (f)=1$ for both second and third generation detectors.

For convenient comparison of $\Omega _{\mathrm{B}}(f)$ with the GW interferometric sensitivity curves discussed above, the noise power spectrum, $S_{\mathrm{n}}(f)$ (in units of $\mathrm{Hz}^{-1}$) over a frequency range $\Delta f$ can be expressed in terms of a detector energy density, $\Omega _{\mathrm{det}} (f)$, over an integration time $T_{\mathrm{int}}$:

\begin{equation}
\Omega _{\mathrm{det}} (f) = \frac{50 \pi^2}{3 H_{0}^2}\frac{S_{\mathrm{n}}(f)}{\Delta f \,T_{\mathrm{int}}\,\gamma (f)}\,f^{3}\,.
 \label{eq_det_closure}
\end{equation}

\noindent The sensitivity curves will be presented for the optimal scenario b), as discussed above for integration times of 1 and 3 years for ET and ALIGO respectively.

\begin{figure}
\begin{center}
\includegraphics*[scale = 0.6]{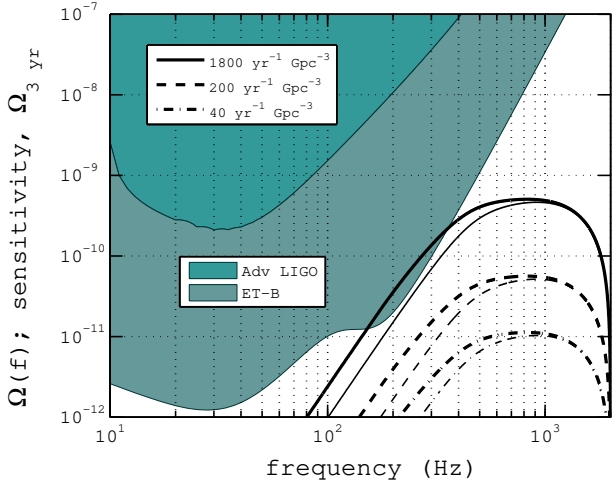}
\end{center}
\caption[]{As for Figure \ref{fig_magnetar_poloidal}, but in the regime where both magnetic and GW emission contribute to the spindown. The triaxial deformations are produced by an internal toroidal field ($B_{t} = 10^{16}$G; $B= 10^{14}$G).}
\label{fig_magnetar_toroidal}
\end{figure}

\section{The GW Background from newly formed magnetars}

\begin{figure}
\begin{center}
\includegraphics*[scale = 0.6]{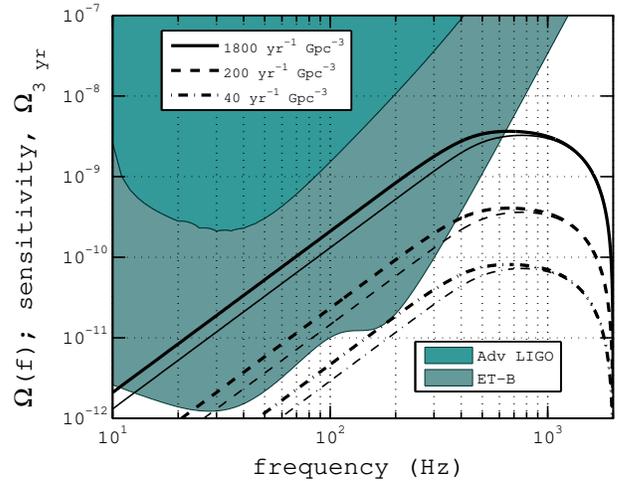}
\end{center}
\caption[]{As for Figure \ref{fig_magnetar_poloidal} but showing upper limits on the GW background spectrum from triaxial deformations in newly born magnetars associated with SL-GRBs. The three curves assume a pure gravitationally induced spindown as given by equation (\ref{eq-Enj_pulsar3}) with $I_{\mathrm{zz}}=10^{45}\mathrm{kg\,m}^2$. We note that in this case, $\Omega$ depends linearly on $I_{\mathrm{zz}}$ and is independent of the ellipticity} \label{fig_magnetar_GW}
\end{figure}

Figures \ref{fig_magnetar_poloidal} to \ref{fig_magnetar_GW} show the function $\Omega_{\mathrm{B}}(f_{\rm{obs}})$ from triaxial deformations in magnetars associated with SL-GRBs for the
three mechanisms discussed in section \ref{sec_GWmagnetars}. Curves are displayed at the rates $(r_{\mathrm{U}},r_{\mathrm{P}},r_{\mathrm{L}})$ and metallicity cutoffs $\epsilon = 0.1$ (thick lines) and $\epsilon = 0.5$ (thin lines). Each plot also includes the detector sensitivities of ALIGO and ET-B assuming 3 years of integration.

Figure \ref{fig_magnetar_poloidal} shows the function $\Omega_{\mathrm{B}}(f_{\rm{obs}})$ assuming that the internal magnetic field is purely poloidal and matches to the external dipole field. For this case the GW emission is negligible in comparison with the magnetic torque and the gravitational signal increases as $f^{4}$ until a maximum at around 840 Hz for rate $r_{\mathrm{U}}$. We see that the signal is outside the sensitivity of ET, and even at the most optimistic rates this background would not be detected within a reasonable integration time. For higher values of $B$ or ellipticity and smaller values of $I_{\mathrm{zz}}$, the amplitude increases until the GW emission dominates ($\Omega \sim f^{2}$) at large frequencies (see fig. 5), eventually reaching a saturation regime (fig. 6).

Figure \ref{fig_magnetar_toroidal} displays the function $\Omega_{\mathrm{B}}(f_{\rm{obs}})$ assuming that internal toroidal fields contribute towards a prolate distortion. For this case both GW and magnetic dipole emissions contribute to the spindown. We see that for the upper rate $r_{\mathrm{U}}$ and $\epsilon = 0.1$ the signal peaks at $\Omega_{\mathrm{B}}(f_{\rm{obs}})\sim 5 \times 10^{-10}$ at around 830 Hz. Even at this upper rate only a small part of this signal is within the sensitivity of ET-B.

\begin{table}
\label{table_mag1}
\begin{centering}
\begin{tabular}[scale=1.0]{lllll}
\hline
\hline
Emission      &ALIGO         &ET-B     &ET-C              \\
Mechanism     &         &     &              \\
  \hline
\hline
\noindent Poloidal      & $\hspace{-1mm}1\times 10^{-5}$    &  0.08                  & 0.03  \\
\noindent field            &$\hspace{-1mm}(6.1\times 10^{-4})$    & \hspace{-1mm}  (0.07)  &\hspace{-1mm}  (0.1) \\
\hline
\noindent Toroidal       & $\hspace{-1mm}2.5\times 10^{-5}$    &  0.2                  & 0.07  \\
\noindent field           &$\hspace{-1mm}(1.5\times 10^{-3})$    & \hspace{-1mm}  (0.5)  &\hspace{-1mm}  (0.2) \\
  \hline
\noindent Pure GW        & 0.006     &  4.5   & 1.7  \\
\noindent Spindown         &\hspace{-1mm} (0.04)    & \hspace{-1mm}  (12.0)  &\hspace{-1mm}  (4.6) \\

\hline
\hline
\end{tabular}
\caption[]{The SNRs obtained through cross-correlation for a GW background of triaxially deformed newly born magnetars associated with SL-GRBs for an event rate of $r_{\mathrm{P}}= 200\hspace{0.5mm} \mathrm{Gpc}^{-3}\mathrm{yr}^{-1}$. Values for the other rates considered in this study,
($r_{\mathrm{U}};r_{\mathrm{L}})= (1800;40), \mathrm{Gpc}^{-3}\mathrm{yr}^{-1}$, can be obtained through the ratios (200/1800;\hspace{0.5mm}200/40). A metallicity cutoff of $\epsilon = 0.1$ is assumed for SL-GRBs. These tabulated values assume 3 years of integration by cross-correlation of data from two detectors. We also assume the overlap reduction functions described in section 5 -- values obtained by optimally orientated and co-located detectors are shown in parentheses. The SNRs obtained from the two emission scenarios considered in section 3 are shown: distortions induced by poloidal fields and toroidal fields; in addition we show upper limits that assume spindown is solely gravitational. We note that the pure GW upper limit could increase up to 3 times the values shown above, as $I_{\mathrm{zz}}$ can be 1--3 times the canonical value used in this study \citep{Ruderman72}.\label{table_magGWBG_1}}
\end{centering}
\end{table}

In Figure \ref{fig_magnetar_GW} we show the upper limit in which spindown is purely gravitational (from equation \ref{eq-Enj_pulsar3}). We see that the background signal increasing with $f^{2}$ and reaching a maximum $\Omega_{\mathrm{B}}(f_{\rm{obs}})\sim 4 \times 10^{-8}$ at around 660 Hz for rate $r_{\mathrm{U}}$. In this case, $\Omega$ depends linearly on $I_{\mathrm{zz}}$ and is independent of the ellipticity.

As illustrated by the thin lines for each curve, which represent $\epsilon = 0.5$, we see that a more relaxed metallicity cutoff results in a smaller contribution below the peak, and hence, a lower SNR, as will be shown below.

Tables 3 and 4 display the SNRs for the three AGBs considered in this section for metallicity cutoffs of $\epsilon=0.1$ and 0.5 respectively. Estimates are shown for $r_{\mathrm{P}}= 200\hspace{0.5mm} \mathrm{Gpc}^{-3}\mathrm{yr}^{-1}$. Values for the rates
($r_{\mathrm{U}};r_{\mathrm{L}})= (1800;40)\, \mathrm{Gpc}^{-3}\mathrm{yr}^{-1}$ can be obtained through the ratios (200/1800;\hspace{0.5mm}200/40). For a GW background produced by toroidal field induced distortions, as illustrated in Fig.\ref{fig_magnetar_toroidal}, at the upper rate $r_{\mathrm{U}}$ a small part of this signal is within the sensitivity of ET-B. This produces a SNR of 1.8 for 3 years of integration. Optimistically, for an upper limit AGB from pure GW spindown, the SNRs confirm that detection would require a third generation detector. Assuming a metallicity cutoff of $\epsilon = 0.1$, we find that source rates of $(133;349)\,\mathrm{Gpc}^{-3}\mathrm{yr}^{-1}$ would result in detection by (ET-B, ET-C) within 3 years at a SNR of 3. A more relaxed metallicity cutoff, $\epsilon = 0.5$, increases the corresponding rates required for detection to $(210;552)\,\mathrm{Gpc}^{-3}\mathrm{yr}^{-1}$. As shown by Table 4 we see that ET-C, which is optimised for greater sensitivity at low frequency, $\leq 20$ Hz, does not improve on the SNR of ET-B. We note that this final scenario can be regarded only as an upper limit, but we consider it as it could allow ET to place constraints on this source population.

\begin{figure}
\begin{center}
\includegraphics*[bbllx = 75pt,bblly =260pt, bburx = 570pt, bbury =590pt,scale = 0.50]{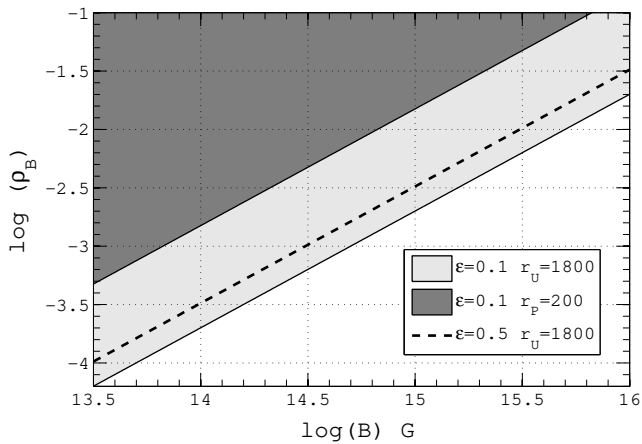}
\end{center}
\caption[]{The $\rho_{B}-B$ plane accessible by ET-B for magnetars associated with SL-GRBs. The shaded zones, set by the product
$(\rho_{B}/10^{-4})\hspace{0.5mm}(10^{14}/B)$, show the parameter space that can be explored for different values of the rate and metallicity $\epsilon$.} \label{fig_constraints}
\end{figure}

We note that in general, independent of the particular mechanism driving the deformation, a NS with an ellipticity $\rho_B$ and an external field $B$, will emit GWs according to equations (4) and (5). The SNR for the corresponding GW background signal will therefore depend on the combination $\rho_{B}/B$.
Treating $\rho_{B}$ and $B$ as independent parameters, and assuming that the NS is born in association with SL-GRB, we can make a statement on detectability that is independent of the actual mechanism causing the ellipticity. This is done in Table 5 by
computing, for each value of the rate ($r_{\mathrm{L}},r_{\mathrm{P}},r_{\mathrm{U}}$) and of the metallicity cutoff $\epsilon$, the minimum $ (\rho_{B}/10^{-4})\hspace{0.5mm}(10^{14}/B)$ required to have a detection with a given detector configuration. The full parameter space is illustrated in Figure \ref{fig_constraints}, for which the product $ (\rho_{B}/10^{-4})\hspace{0.5mm}(10^{14}/B)$, shown by the diagonal lines in the $\rho_{B}-B$ plane, divides the plot into detectibility zones (shown by the legend). We see that for a rate of $r_{\mathrm{L}}$ even pure GW emission is out of reach. However, with $r_{\mathrm{P}}$ we could access extreme values and with $r_{\mathrm{U}}$ a large part of parameter space is detectable.

%
%
%
%

In the next section we will consider the background signal from secular instabilities
which occur on a shorter timescale than the emissions considered in this section, $\sim 1000$\hspace{0.5mm}s, corresponding with the X-ray plateaus observed in some LGRBs. For this signal some analysis of the $DC$ will be important.

\begin{table}
\label{table_mag1}
\begin{centering}
\begin{tabular}[scale=1.0]{lllll}
\hline
\hline
Emission      &ALIGO         &ET-B     &ET-C              \\
Mechanism     &         &     &              \\
  \hline
  \hline
\noindent Poloidal      & $\hspace{-1mm}4.2\times 10^{-6}$    &  0.04                  & 0.01  \\
\noindent field            &$\hspace{-1mm}(3.4\times 10^{-4})$    & \hspace{-1mm}  (0.1)  &\hspace{-1mm}  (0.04) \\
\hline
\noindent Toroidal       & $\hspace{-1mm}1.1\times 10^{-5}$    &  0.1                  & 0.04  \\
\noindent field           &$\hspace{-1mm}(8.5\times 10^{-4})$    & \hspace{-1mm}  (0.3)  &\hspace{-1mm}  (0.1) \\
\hline
\noindent Pure GW        & 0.004     &  2.8   & 1.3  \\
\noindent Spindown         &\hspace{-1mm} (0.03)    & \hspace{-1mm}  (7.6)  &\hspace{-1mm}  (3.4) \\
\hline
\hline
\end{tabular}
\caption[]{As for Table 4 but with a metallicity cutoff of $\epsilon = 0.5$  for SL-GRBs.}
\end{centering}
\label{table_mag2}
\end{table}

\begin{table}
\label{table_constraints}
\centering
\begin{tabular}[scale=1.0]{lcc}
  \hline
  \hline
  Rate  & $\epsilon=0.1$ & $\epsilon=0.5$\\
  ($\mathrm{Gpc}^{-3}\mathrm{yr}^{-1}$) &&\\
    \hline
  \hline
   40 & - & - \\
   200 & 15 & pure GW \\
   1800 & 2 & 3 \\
  \hline
  \hline
\end{tabular}
\caption[]{The minimal value of the product $(\rho_{B}/10^{-4})\hspace{0.5mm}(B/10^{14})^{-1}$ required to obtain a SNR of 3 with ET-B over 3 years of observation, for the 3 rates considered in this paper and for metallicity cutoffs of $\epsilon=0.1$ and $\epsilon=0.5$. An omitted value implies that a detection is not expected.}
 \end{table}


\section{The GW Background from secular bar-mode instabilities}

\subsection{The GW background spectrum}
\label{Magnetar_AGB}

The SNRs estimated for a background signal from secular bar mode instabilities in SL-GRBs are
shown in Tables \ref{table_secular01} and \ref{table_secular05} for the rates $(r_{\mathrm{U}}, r_{\mathrm{P}}, r_{\mathrm{L}})$. The estimates indicate that ALIGO will require 3 yrs of integration by optimally orientated and co-located detectors to reach a SNR of 1 for the optimistic rate $r_{\mathrm{U}}$. For ET-B however, this signal can potentially be detected with a SNR $ \geq 3$ in the more conservative hypothesis for SL-GRB rates $r_{\mathrm{L}}$ and detector performances.

We stress here that these estimates are based on two main assumptions:

\begin{enumerate}
  \item that at least 40\% of SL-GRBs are associated with magnetar progenitors undergoing a secular-bar mode instability;
  \item that the magnetar's parameters are those adopted to calculate the single-source spectrum shown in Fig. 1.
\end{enumerate}

Regarding (i), we note that there is significant uncertainty in how often SL-GRBs could be associated with a secularly unstable magnetar progenitor. This, in turn, implies a large uncertainty on our rate estimates. To address this problem, we have chosen a wide range of values.

We note again that our lower rate $r_{\mathrm{L}}$ accounts for the fraction of LGRBs showing X-ray plateaus in the \emph{Swift} Era -- around 40\% \citep{Evans_09}. We suggest that $r_{\mathrm{L}}$ is a reasonable estimate also in view of the uncertainty (ii) underlined above. In fact, the parameter values adopted by \citet{Corsi:2009} aimed at
explaining the typical case of a $\sim 1000$\hspace{0.5mm}s duration plateau observed in a LGRB with an energy release similar to those of SL-GRBs. It is thus more conservative to assume that the spectrum shown in Fig. \ref{fig_dedf_secular} would be realized only in a fraction of SL-GRBs similar to the one of LGRBs showing a plateau $(\sim 40\%)$. We note that fraction could be higher in SL-GRBs, since dipole energy injection from a magnetar can more easily cause visible plateaus on less energetic GRBs. However, it is not yet clear whether or not X-ray plateaus are always caused by a magnetar -- see e.g. \citet{Panaitescu_08} for an alternative explanation. In the light of these uncertainties, we consider $r_{\mathrm{L}}$ as a safer estimate.

We calculate that rates of $(48,80)\,\mathrm{Gpc}^{-3}\mathrm{yr}^{-1}$ for (ET-B, ET-C) are required to achieve a SNR of 3 for 1 year of integration by separated detectors. For our conservative rate $r_{\mathrm{L}}$, (ET-B, ET-C) will require (1.4, 4) yrs of integration.

\begin{figure}
\begin{center}
\includegraphics*[scale = 0.6]{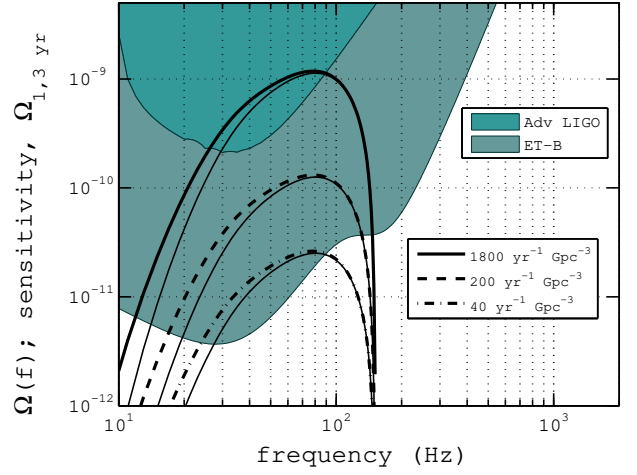}
\end{center}
\caption[]{Upper limits on the GW background spectrum from secularly unstable magnetars associated with under-luminous GRBs for the rates $(r_{\mathrm{U}},r_{\mathrm{P}},r_{\mathrm{L}})= (1800,200,40)\, \mathrm{Gpc}^{-3}\mathrm{yr}^{-1}$. The sensitivity curves of (ALIGO, ET-B) are shown in terms of $\Omega_{\mathrm{det}}(f)$ and assume (3, 1)\hspace{0.5mm}yrs of integration and optimally orientated and co-located detectors. The thin curves show the background estimates for a more relaxed metallicity dependence, $\epsilon = 0.5$.}\label{fig_secular_ET}
\end{figure}

Figure \ref{fig_secular_ET} shows the quantity $\Omega_{\mathrm{B}}(f_{\rm{obs}})$ for the rates $(r_{\mathrm{U}},r_{\mathrm{P}},r_{\mathrm{L}})$, in comparison with the sensitivity curves for second and third generation GW interferometric detectors represented by ALIGO and ET-B. The stochastic background signal has a frequency bandwidth 5--150\hspace{0.5mm}Hz with a peak of $\Omega_{\mathrm{B}}(f_{\rm{obs}}) \sim 10^{-9}$ at around 80 Hz. The thin lines in the figure show the function $\Omega_{\mathrm{B}}(f_{\rm{obs}})$ assuming a more relaxed allowance for metallicity, $\epsilon = 0.5$. This illustrates once again how a lower metallicity cutoff results in a greater contribution of $\Omega_{\mathrm{B}}(f_{\rm{obs}})$ at lower frequency. In Appendix A we will further discuss the effect of cosmic metallicity on the GW background signal.

\begin{figure}
\begin{center}
\includegraphics*[scale = 0.6]{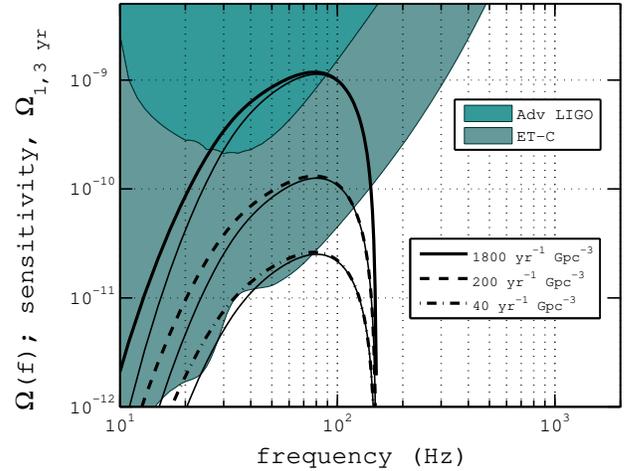}
\end{center}
\caption[]{As for Figure \ref{fig_secular_ET}, but showing an alternative configuration for the Einstein Telescope, ET-C.} \label{fig_secular_ETC}
\end{figure}

Figure \ref{fig_secular_ETC} compares the function $\Omega_{\mathrm{B}}(f_{\rm{obs}})$ with ET-C. In comparison with Fig. \ref{fig_secular_ET}, we see that at the conservative rate, $r_{\mathrm{L}}$, the only significant contribution from this signal is from $\gtrsim 50$ Hz. As can be seen by a comparison of Tables 5 and 6, this results in a SNR of around a factor of 2 less. At the plausible rate $r_{P}$ there is no contribution below $\sim 15$ Hz. Therefore, for the most probable rate estimates, ET-C could still pursue a primordial GW background signal at its most sensitive frequency bandwidth.

For ET-B this signal occurs in the most sensitive frequency regime; a largely continuous signal could therefore mask any primordial GW background signal with $\Omega(f) \leq  10^{-9}$ within the bandwidth 20--100\hspace{0.5mm}Hz. A $DC$ analysis will indicate what proportion of sources will contribute to a continuous signal. This will have implications for both for stochastic searches and for single detections by ET, for which the detection horizon may enter the $DC\geq 10$ confusion limited regime. We shall investigate this in the next section.

\begin{table}
\begin{centering}
\begin{tabular}[scale=1.0]{cccc}
\hline
\hline
Rate    &ALIGO         &ET-B     &ET-C              \\
$\mathrm{Gpc}^{-3}\mathrm{yr}^{-1}$ & &\\
  \hline
  \hline
1800        &  1 (4)   & 113 (300)  & 67 (178)\\
200        &  0.1 (0.4)   & 13 (33)  & 8 (20)\\
40         & 0.02 (0.08)   & 3 (7) & 2 (4)\\
\hline
\hline
\end{tabular}
\caption[]{The SNRs achievable through cross-correlation of 3 yrs of data by ALIGO and 1 yr by ET-B or ET-C, for a GW background resulting from secularly unstable magnetars associated with SL-GRBs. Values are shown for the rates $(r_{\mathrm{U}},r_{\mathrm{P}},r_{\mathrm{L}})= (1800,200,40)\, \mathrm{Gpc}^{-3}\mathrm{yr}^{-1}$ and assume a metallicity cutoff $\epsilon = 0.1$. The SNR estimates assume the overlap reduction functions described in section \ref{section_detection} -- values obtainable by optimally orientated and co-located detectors are shown in parentheses. \label{table_secular01}}
\end{centering}
\end{table}

\begin{table}
\begin{centering}
\begin{tabular}[scale=1.0]{cccc}
\hline
\hline
Rate    &ALIGO         &ET-B     &ET-C              \\
$\mathrm{Gpc}^{-3}\mathrm{yr}^{-1}$ & &\\
  \hline
  \hline
1800        &  0.7 (3)   & 96 (257)  & 55 (148)\\
200        &  0.09 (0.3)   & 11 (29)  & 6 (16)\\
40         & 0.02 (0.07)   & 2 (6) & 1 (3)\\
\hline
\hline
\end{tabular}
\caption[]{As for Table \ref{table_secular01} but with a metallicity cutoff of $\epsilon = 0.5$  for SL-GRBs.\label{table_secular05}}
\end{centering}
\end{table}


\subsection{The duty cycle}

For a GW from triaxial deformations in magnetars associated with SL-GRBs we find that for rates $(r_{\mathrm{U}},r_{\mathrm{P}},r_{\mathrm{L}})= (1800,200,40)\,\mathrm{Gpc}^{-3}\mathrm{yr}^{-1}$ sources outside a volume defined by $z\sim (0.07, 0.1, 0.2)$ contribute to a continuous signal. This is the result of a long duration, $\tau \sim 10^{6}$\hspace{0.5mm}s in equation \ref{eq_DC} \citep{stella_05aa}.

\begin{figure}
\begin{center}
\includegraphics[scale = 0.55]{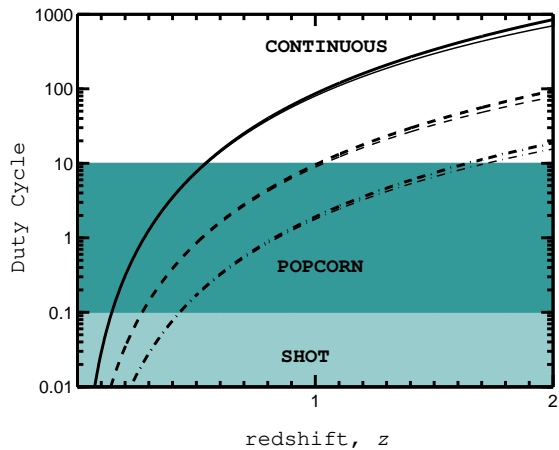}
\end{center}
\caption[]{Duty cycle as a function of redshift for a GW background from secularly unstable magnetars in SL-GRBs occuring at rates of (40, 200, 1800) $\mathrm{Gpc}^{-3}\mathrm{yr}^{-1}$. The shaded areas show three zones of an AGB corresponding to different regimes of $DC$: continuous ($DC \geq 10$), popcorn ($0.1\leq DC < 10$) and shot noise ($DC \leq 0.1$) \citep{coward_regimbau_06}. We see that sources beyond $(z = 0.5, 1.0, 1.6)$ contribute to a continuous signal.}
\label{fig_dc}
\end{figure}

Figure \ref{fig_dc} shows the duty cycle as a function of redshift for a GW background resulting from secularly unstable magnetars occurring in SL-GRBs. For $\tau$, we use a value of 1000\hspace{0.5mm}s -- this approximates to the typical duration of an X-ray plateau for a GRB.
The plot shows that as the rate decreases, the continuous contribution to the background is from sources at greater distances. For the rates $(r_{\mathrm{U}},r_{\mathrm{P}},r_{\mathrm{L}})= (1800,200,40)\,\mathrm{Gpc}^{-3}\mathrm{yr}^{-1}$ we find a $DC \geq 10$ is reached at around $z\sim (0.5, 1.0, 1.6)$. Therefore sources outside volumes defined by these $z$ values will contribute to a continuous background signal. Curves for a more relaxed cutoff $\epsilon = 0.5$ are shown by the thinner lines. Referring to Figure \ref{fig_drdz}, we see that the effect of metallicity dependence is small within  $z \sim 1$. This reflected in the curves of Figure \ref{fig_dc}.

The optimal and isotropic (orientation averaged) horizon distances \citep[see][for further definitions]{Regimbau:2009rk,Dalal_SGRB_Sirens_06} are greatest for the ET-B detector, at distances of $z = (0.2, 0.12$), both less than the redshift range in which the signal becomes continuous. Thus, a confusion-limited background will not affect the resolution of individual sources.

\begin{figure}
\begin{center}
\includegraphics*[scale = 0.6]{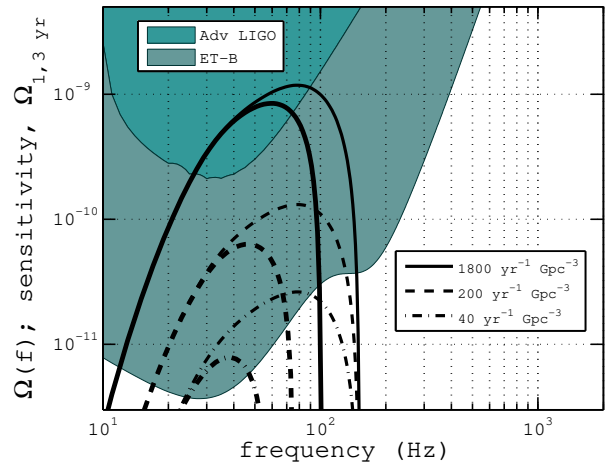}
\end{center}
\caption[]{As for Figure \ref{fig_secular_ET}, but with thick lines showing only the continuous contribution ($DC \geq 10$) from sources ($z \geq 0.5$, $z \geq 1.0$, $z \geq 1.6$) for the rate estimates $(r_{\mathrm{U}},r_{\mathrm{P}},r_{\mathrm{L}})$.} \label{fig_secular_DC}
\end{figure}

Figure \ref{fig_secular_DC} shows again the results of Fig. \ref{fig_secular_ET}, but with thick curves to show the continuous contributions to the AGB signal from sources at $z$ greater than $(0.5, 1.0, 1.6)$. We see that even if the popcorn and shot components can be identified, the continuous part of the AGB could mask a primordial GW background signal in the most sensitive frequency regime, around 20 -- 30\hspace{0.5mm}Hz, of ET-B. As rates increase, so to does the continuous proportion of the AGB signal.

\section{Conclusions}

In this paper we have assessed the GW detection prospects for the background signals associated with SL-GRBs, assuming that the central engines of a significant proportion of these bursts are provided by newly born magnetars. We have considered two plausible GW single-source emission mechanisms: a) the deformation-induced GW emission from a newly born magnetar b) the onset of a secular bar-mode instability. The latter mechanism would correspond well with the long lived shallow plateau observed in the X-ray afterglows of many GRBs.

We have calculated the GW background spectra of each of the mechanisms by employing appropriate models for each. We account for GRBs preference towards low-metallicity environments by using a source rate history model that allows for cosmic metallicity evolution. We assume both a low metallicity cutoff defined by $\epsilon = 0.1$ and a more relaxed cutoff, $\epsilon = 0.5$.

Our results for the deformation-induced GW emission from a newly born magnetars are more pessimistic than those presented by \citet{regimbau_astBG_08}. This is due to the fact that whilst they considered the emission from the population of magnetars assuming they represented 10\% of the population of newborn neutron stars (a rate  of around $(3-15)\times10^{3}\,\mathrm{Gpc}^{-3}\mathrm{yr}^{-1}$), we consider only magnetars associated with SL-GRBs.

For an AGB from triaxial emissions in newly formed magnetars associated with SL-GRBs, for an upper limit case in which emission is purely from GW emission, rates of $(52,137)\,\mathrm{Gpc}^{-3}\mathrm{yr}^{-1}$ will be required for detection at a SNR of 3 within 1 year by (ET-B, ET-C). For a AGB resulting from from toroidal fields, an upper rate $r_{\mathrm{U}}=1800\,\mathrm{Gpc}^{-3}\mathrm{yr}^{-1}$ would produce a SNR of 1.8 after 3 years of integration by ET-B. We find however that rates above $200\hspace{0.5mm} \mathrm{Gpc}^{-3}\mathrm{yr}^{-1}$ would enable ET-B to explore the $\rho_{B}-B$ parameter space of the magnetar population considered in this study.

In terms of detectability, we find that an AGB resulting from the onset of a secular instability is a more optimistic scenario. We note however, that this is highly dependent on rate of occurrence of this instability. Using the single-source GW emission model of \citet{Corsi:2009}, we find that event rates of $(48,80)\,\mathrm{Gpc}^{-3}\mathrm{yr}^{-1}$ are sufficient to produce a detectable signal for ET (ET-B, ET-C) with  SNR of 3 for 1 yr of observation. For ALIGO, detection within 3 years would require the upper limit rate estimate $r_{\mathrm{U}}$, combined with a pair of optimally orientated and co-located detectors. We note that observations of a larger number of SL-GRBs \citep[e.g. by future satellites like Janus or
EXIST;][]{2009astro2010S.284S,Imerito_08}, will help in reducing the uncertainties on their local rate estimates, thus clarifying the prospects of detectability of an associated GW background.

We find that this signal could potentially mask a primordial GW background signal. Analysis of the $DC$ for the background signal from secularly unstable magnetars showed that even at a conservative rate estimate $r_{\mathrm{L}}$, a significant proportion of this signal would be continuous. As highlighted in Fig. \ref{fig_secular_DC}, this would occur in the most sensitivity bandwidth of ET-B.
Depending on the rate of occurrence, both mechanisms could produce GW backgrounds that could mask a primordial GW background signal of order $\Omega _B (f) \sim 5 \times 10^{-11}$ in the frequency regime 10\,--\,50\hspace{0.5mm}Hz of ET-B and ET-C. This would pose a particular problem for the former, as it is the most sensitive bandwidth of this detector.

%

The AGB may form a composite signal with an AGB from NS/NS inspirals, which is expected across a bandwidth of 10 -- 800\hspace{0.5mm}Hz with increasing $\Omega _B$ \citep{regimbau_NSBG_ApJ_06}. As the latter background will peak at $\sim 1000$ Hz, detecting the higher frequency component may enable the two AGBs to be disentangled.

To calculate SNRs we have applied equation (\ref{snr}), which assumes the background signals can be detected through cross-correlation. We have chosen to adopt this convention for easy comparison with other astrophysical GW background estimations, many of which will contain popcorn or shot noise components. In practice, to detect the non-continuous components of a GW background some other strategy will be required. Given the long duration and quasi-periodicity of the signal from secular instabilities, ET could detect a significant number of the shot noise events through matched filtering. This provides an additional means to interrogate the higher energy, shot noise and popcorn components of the background signal which result from the rarer nearby events. A statistical procedure, such as the ``probability event horizon" technique, which extracts the observation time dependence from a population of cosmological transients could be used \citep{coward_PEH_05}. This technique, which has been used to place constraints on the rate density of source populations, could use single detections from the shot noise component to interrogate the temporal dimension of the GW background signal \citep{howell07_APJL,howell07,coward_firstdetection_08,howell:203}.


Although the AGB signals discussed here remain speculative, and the estimated rates are still highly uncertain, detection of the GW emission mechanisms associated with SL-GRB events could yield important payoffs. This could be possible should locally observed SL-GRBs produce the necessary triggers for multi-messenger observations. Coupled with a single-source detection, an AGB signal could provide constraints on the high-$z$ evolution of the highly flux-limited SL-GRB population and would be a valuable probe of source rate and metallicity evolution.

%
%

\appendix
\label{app1}
\section{The effect of cosmic metallicity on a GW background signal}
\begin{figure}
\begin{center}
\includegraphics*[scale = 0.6]{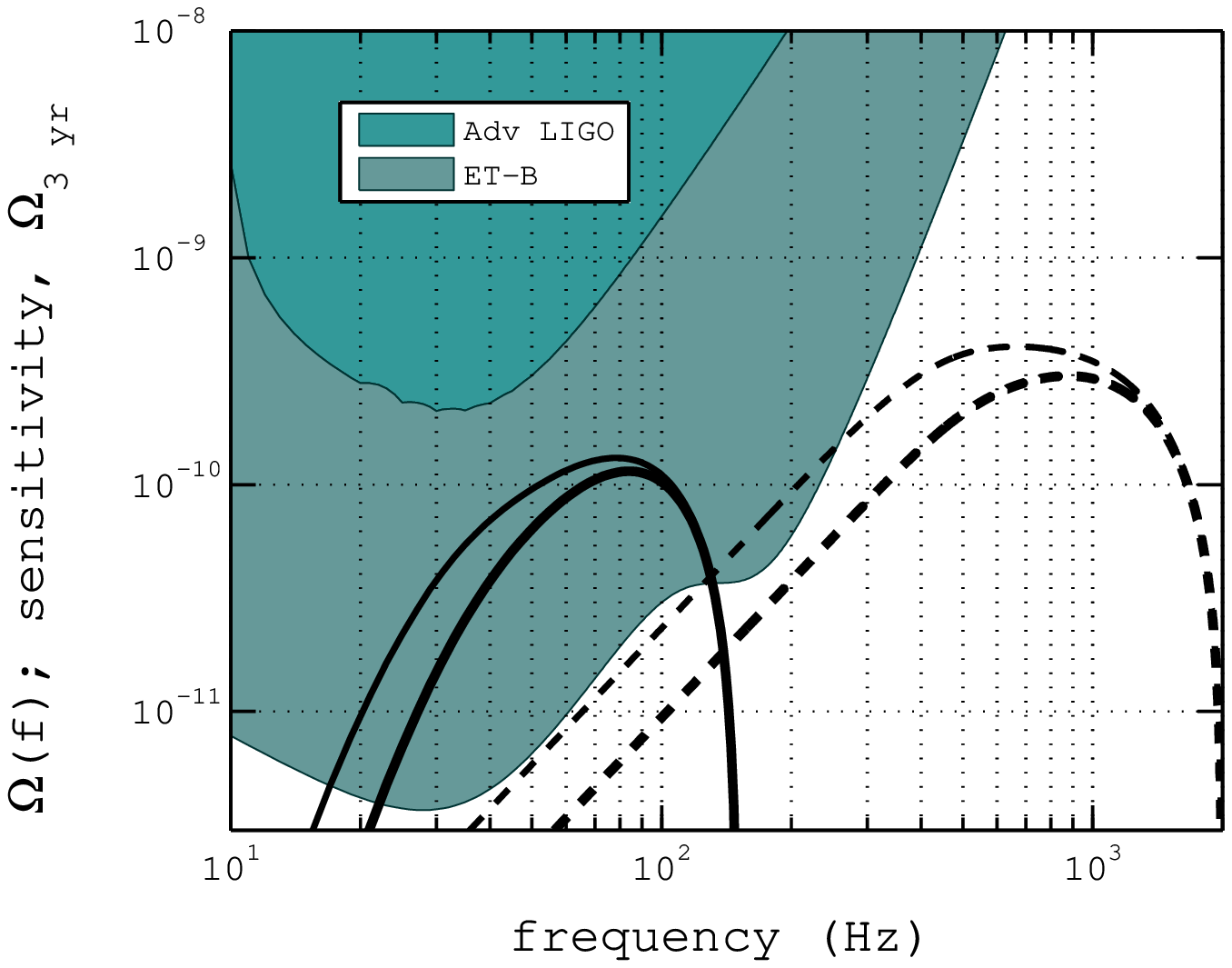}
\end{center}
\caption[]{To illustrate the effect of metallicity dependence on the GW background signal we reproduce the curves of Figs\hspace{1mm}6 and \ref{fig_secular_ET} for rate $r_{\mathrm{P}}$. We add thick lines showing a metallicity independent source rate evolution assuming $\epsilon = 1.0$. We see that including a metallicity dependence shifts the background spectrum slightly to lower frequency.} \label{fig_secular_met}
\end{figure}

To illustrate the effect of metallicity dependence on calculations of a GW background signal, in Figure \ref{fig_secular_met} we reproduce the curves of Figs \ref{fig_magnetar_GW} and \ref{fig_secular_ET} for the rate $r_{\mathrm{P}}$. We add thick lines showing a metallicity independent ($\epsilon = 1.0$) source rate evolution. We see that including a metallicity cutoff of $\epsilon = 0.1$, gives a higher signal at lower frequencies. This is a result of the greater contribution from high-$z$ sources illustrated in Fig. \ref{fig_drdz}. As indicated by the results in tables 3 -- 6, this effect increases the SNR estimates.

\section*{Acknowledgments}
We gratefully thank  P. {M{\'e}sz{\'a}ros} for agreeing to provide us with data for the model in \citet{Corsi:2009} of the secular bar-mode instability. We also thank Christian Ott for a careful reading an early initial manuscript and for providing us with important feedback on post-collapse instabilities in core-collapse supernovae. The authors also gratefully acknowledge Vuk Mandic who carefully read the manuscript and made some insightful suggestions as part of the LIGO Scientific Collaboration review. LIGO was constructed by the California Institute of Technology and Massachusetts Institute of Technology with funding from the National Science Foundation. This paper has LIGO Document Number LIGO-P100083.

\bibliographystyle{mn2e} 
\bibliography{magnetar_mnras}
\end{document}